\documentclass[aps,pre,twocolumn,groupedaddress,showpacs]{revtex4-1}
\usepackage{graphicx}
\usepackage{amssymb}
\usepackage{amsmath}
\usepackage{textcomp}
\usepackage{mathtools}
\usepackage{psfrag}

\newcommand{\Ca}{{\rm Ca}}
\newcommand{\Caup}{{\rm Ca}_{\rm up}}
\newcommand{\Calo}{{\rm Ca}_{\rm lo}}
\newcommand{\Cacr}{{\rm Ca}_{\rm cr}}
\newcommand{\M}{{\rm M}}
\newcommand{\keff}{\kappa_{\rm eff}}
\newcommand{\kcr}{\kappa_{\rm eff, cr}}
\newcommand{\mueff}{\mu_{\rm eff}}
\newcommand{\Vdiss}{V_{\rm diss}}
\newcommand{\Ddiss}{D_{\rm diss}}
\newcommand{\dtr}{d_{\rm tr}}
\newcommand{\etal}{et al.}

\psfrag{Ca}[][][1.8]{$\Ca$}
\psfrag{Keff}[][][2.2]{$\keff$}
\psfrag{u,}[][][1.8]{$u$,}
\psfrag{CamCatr}[][][2.0]{$\Ca - \Cacr$}
\psfrag{keffmktr}[][][2.0]{$\keff - \kcr$}

\begin{document}

\title{Two-Phase Flow in Porous Media: Scaling of
Steady-State Effective Permeability}

\author{Morten Gr{\o}va}
\email{Morten.Grova@ntnu.no}

\affiliation{
Department of Physics,
Norwegian University of Science and Technology,
NO-7491 Trondheim,
Norway}

\date{April 19, 2012}

\begin{abstract}
A recent experiment has considered the effective permeability of two-phase flow
of air and a water-glycerol solution under steady-state conditions in a
two-dimensional model porous medium, and found a power law dependence with
respect to capillary number. Running simulations on a two-dimensional network
model similar power law behavior is found, for high viscosity contrast as in the
experiment and also for viscosity matched fluids. Two states are found, one
with stagnant clusters and one without. For the stagnant cluster state, a power
law exponent $0.50$ is found for viscosity matched fluids and $0.54$ for large
viscosity contrast.
\end{abstract}


\maketitle


\section{Introduction}


A system consisting of two immiscible fluids flowing simultaneously through a
porous medium has provided a complex challenge for both scientists and
engineers \cite{dullien1992porous, sahimi1995flow}. An improved understanding
of two-phase flow in porous media is of considerable importance for such
applications as enhanced oil recovery and groundwater management. Also,
attention has recently been directed towards the reduced efficiency of polymer
electrolyte membrane (PEM) fuel cells caused by produced water flooding the
pore space of the PEM's gas diffusion and catalyst layers
\cite{mukherjee2011pore}.


Simulations of two-phase flow in porous media may be done by modeling the
porous medium as a network \cite{blunt1992simulation, aker1998two,
blunt2001flow, algharbi2005dynamic, idowu2010pore} or by simulating fluid
dynamics directly on porous media reconstructions from microstructure images
\cite{ramstad2011relative}. The advantage of the latter approach is a more
realistic description of the pore space. Network modeling, on the other hand,
has the advantage of allowing a large range of simulation parameters on
reasonably large systems, even when the viscous pressure field needs to be
re-solved for each timestep.

Most work on two-phase flow in porous media, both numerical and experimental,
has focused on displacement processes \cite{lenormand1988numerical}. Relative
permeabilities, used in reservoir modeling, can be measured under such
un-steady conditions. Another approach is to maintain constant external control
parameters, such as total flow-rate and average saturation, and run the
simulation until macroscopic quantities such as relative permeabilities only
fluctuate around mean values. These are steady-state values, which in general
will be different from the un-steady values. An open question is whether there
exists some range of control parameters within which the steady-state values
are unique, independent of initial conditions.


Network steady-state simulations have considered various aspects such
as the dynamics of disconnected ganglia of oil, the phase diagram in the regime
where both phases are mobile, the transition where one of the phases becomes
immobilized and diffusion on non-wetting clusters
\cite{dias1986networkPart1, dias1986networkPart2,
constantinides1996network, valavanides1998mechanistic, valavanides2001true,
knudsen2002bulk, knudsen2006two, ramstad2006cluster}.


Experimental work on steady-state with simultaneous flow of both phases has
been done in two dimensions using chamber-throat networks etched in glass
\cite{avraam1994steady, avraam1995generalized, avraam1995flow, avraam1999flow,
tsakiroglou2007transient} and Hele-Shaw cells \cite{tallakstad2009prl,
tallakstad2009pre}. A recent experiment performed with three-dimensional bead
packings \cite{rassi2011nuclear} is also of interest to this work. Recently,
microfluidic devices have been used as model porous media, achieving
unprecedented control over the pore-scale geometry, but so far they have only
been used to study drainage \cite{cottin2010drainage}.


When studying steady-state two parameters are usually kept under control. Flow
is then maintained until all measurable quantities only fluctuate around some
mean value. One control parameter is the capillary number $\Ca$, a
dimensionless quantity which parametrizes the relative strength of viscous and
capillary pressure drops. It is proportional to the total flow-rate. The other
parameter which controls the steady-state may be either the non-wetting
saturation $S$ or the fractional flow of the non-wetting phase $F$. The
pressure drop required to drive flow may be measured, as well as relative
permeabilities. The sum of relative permeabilites is the effective permeability
$\keff$. This is typically found to be less than unity under steady-state
conditions, which signifies that the mixture of two phases results in a lower
permeability than if only a single phase is present.

Tallakstad~\etal ~\cite{tallakstad2009prl, tallakstad2009pre} have reported
that the pressure drop $\Delta P$ driving flow of air and a water-glycerol
solution under steady-state conditions in a two-dimensional model porous medium
shows a power law dependence with respect to the capillary number $\Ca$,
\begin{equation}
\label{eq:beta}
  \Delta P \sim \Ca^{\beta} .
\end{equation}
They keep fractional flow fixed at $F=7/15$, due to injecting the two phases
through 15 syringes of equal flow-rates, 7 of which inject air. In the following
this will be referred to as the Oslo experiment.

Rassi~\etal ~\cite{rassi2011nuclear} have also reported a power law similar to
Eq.~\ref{eq:beta}. Their experiment uses three-dimensional bead packings
partially saturated with air under flowing water conditions. The partially
saturated state is achieved by initial simultaneous flow of both two phases,
combining them at a tee junction beneath the bead pack. In the following this
experiment will be referred to as the Bozeman experiment.

Using a two-dimensional network simulation this work investigates whether power
law behavior similar to Eq.~\ref{eq:beta} is found when saturation is kept
fixed, and whether it extends beyond the limit of large viscosity ratios used
in the experiments. A preliminary report has been published in
\cite{grova2011two}.


\section{The model}


\subsection{General}

The porous medium is modeled as a network of connected capillaries. A
capillary corresponds to the throat connecting two pores, within which
one or more menisci may be present. With regards to the pressure drop
across a meniscus the throats are modeled as having an hour glass
shape, while the flow equations are solved with permeabilities
corresponding to cylindrical tubes. This is a matter of convenience,
and reflects an interest in the network-scale dynamics rather than a
realistic description at the pore scale.

The network considered in this work is a square lattice inclined at
45{\textdegree} relative to the main direction of flow. Boundary
conditions are bi-periodic, such that the network may be mapped onto a
torus. Disorder is introduced by assigning random radii $r$ to
throats, from a flat distribution between 10\% and 40\% of the throat
length. All throats are assumed to be equally long, with length
$\ell$.

A modified version of the Young-Laplace relation is used to obtain
the capillary pressure drop
\begin{equation}
  p_c(x) = \frac{2\sigma}{r} \left( 1-\cos(\frac{2\pi x}{\ell}) \right) ,
\end{equation}
where $x$ is the position of the meniscus in the interval $[0,\ell]$
and $\sigma$ is the surface tension.

Disregarding wetting films and considering flow to be laminar, the
Hagen-Poiseuille permeability for cylindrical tubes gives the flow-rate
$q$ by the Washburn equation,
\begin{equation}
\label{eq:washburn}
  q = -\frac{\pi r^4}{8\ell \mu} \left( \Delta p + \sum p_c \right) ,
\end{equation}
where $\mu$ is the viscosity of the phase contained in the throat and
$\Delta p$ is the difference between the pressures of the two nodes
connected by the throat.
If the viscosities of the two phases are different, the viscosity $\mu$
of Eq.~\ref{eq:washburn} is replaced by the volume-weighted effective
viscosity averaged over the length of the throat.
The two phases are considered to be incompressible.

Flow is driven by a pressure drop applied across a cut through the
network. The cut is invisible to the local flow field, so it does
not cause any boundary effects.
A constant volumetric flow-rate $Q$ is achieved by solving the pressure
field for two applied pressure drops $\Delta P$ and extrapolating linearly
to the pressure drop which gives the desired $Q$ using the relationship
\begin{equation}
  Q = Q_0 + Q_c = -\frac{\kappa_0}{\mueff} \frac{\phi A}{L_y} \Delta P + Q_c ,
\end{equation}
where the first term is the single-phase flow-rate $Q_0$ corresponding to
Darcy's law and $Q_c$ is a flow-rate induced by the capillary pressure drops,
i.e., it is the flow-rate the system would have if the external pressure drop
was set to zero.
Under steady-state conditions $Q_c$ is always in the opposite direction of
$Q$, i.e., the net effect of the capillary pressure drops is to oppose flow.
In the Darcy term $Q_0$, $L_y$ is the length of the sample in the direction
of flow, $A$ is the cross-sectional area of the sample and $\phi$ is its porosity.
$\kappa_0$ is the single-phase absolute permeability of the porous medium and
$\mueff$ is the volume-weighted effective viscosity averaged over the entire system.

After solving the pressure field, the simulation is iterated forward
in time. The throat with largest flow-rate is identified and the
timestep is set by allowing any menisci in this throat to advance 10\%
of the throat length. Most throats will have flow-rates orders of
magnitude slower than this, giving a smooth motion of menisci through
the hour glass shaped throats. The 10\% rule is chosen by trial and
error as being as large as possible without affecting the results of
the simulation. Timestepping is done by the Euler method, which is
only first order accurate. The numerical error may be estimated by
comparing the difference between driving power and heat dissipation,
which should average out to zero in the steady-state. It is found that
the error is negligible for $\Ca > 10^{-3}$, but grows to a few
percent for $\Ca = 10^{-4}$. For $\Ca$ lower than this the Euler
method is therefore not reliable.

Since the network is a closed system (mapped onto a torus) the total
volume of the two phases is conserved, i.e., any given run of the
simulation maintains a constant saturation $S$ which is determined by
the initialization conditions.

When a meniscus reaches the end of a throat, new menisci are created
in throats flowing out of the connecting pore. In order to prevent the
number of menisci from growing without bonds some scheme is needed to
merge nearby pairs of menisci. Menisci within a certain distance from
the ends of throats may be merged, mimicking fluid coalescence withing
the pore volumes. The closest pair of menisci may be merged whenever
some maximum allowed number of menisci within a throat is reached. A
minimum separation between menisci may also be enforced, merging pairs
of menisci when they come too close. Whenever pairs of menisci are
merged, this is done in such a way that the merging does not
contribute to fractional flow. All these schemes require some
coalescence parameter. All have been implemented and tested. It is
found that while fractional flow depends on the merging scheme and
coalescence parameter, the power law exponent does not. Only the range
of validity of the power law is affected, as it ceases to be valid
when fractional flow becomes either 0 or 1.

For the results given here, pairs of menisci less than 10\% of the
length of a throat apart are merged.

A more realistic description would need to carefully consider
mechanisms for snap-off and coalescence. As a detailed description of
flow at the pore scale is not the primary interest of this work, the
simpler approach is considered sufficient.

\begin{figure*}[t]
\begin{center}
  \scalebox{0.32}{\includegraphics{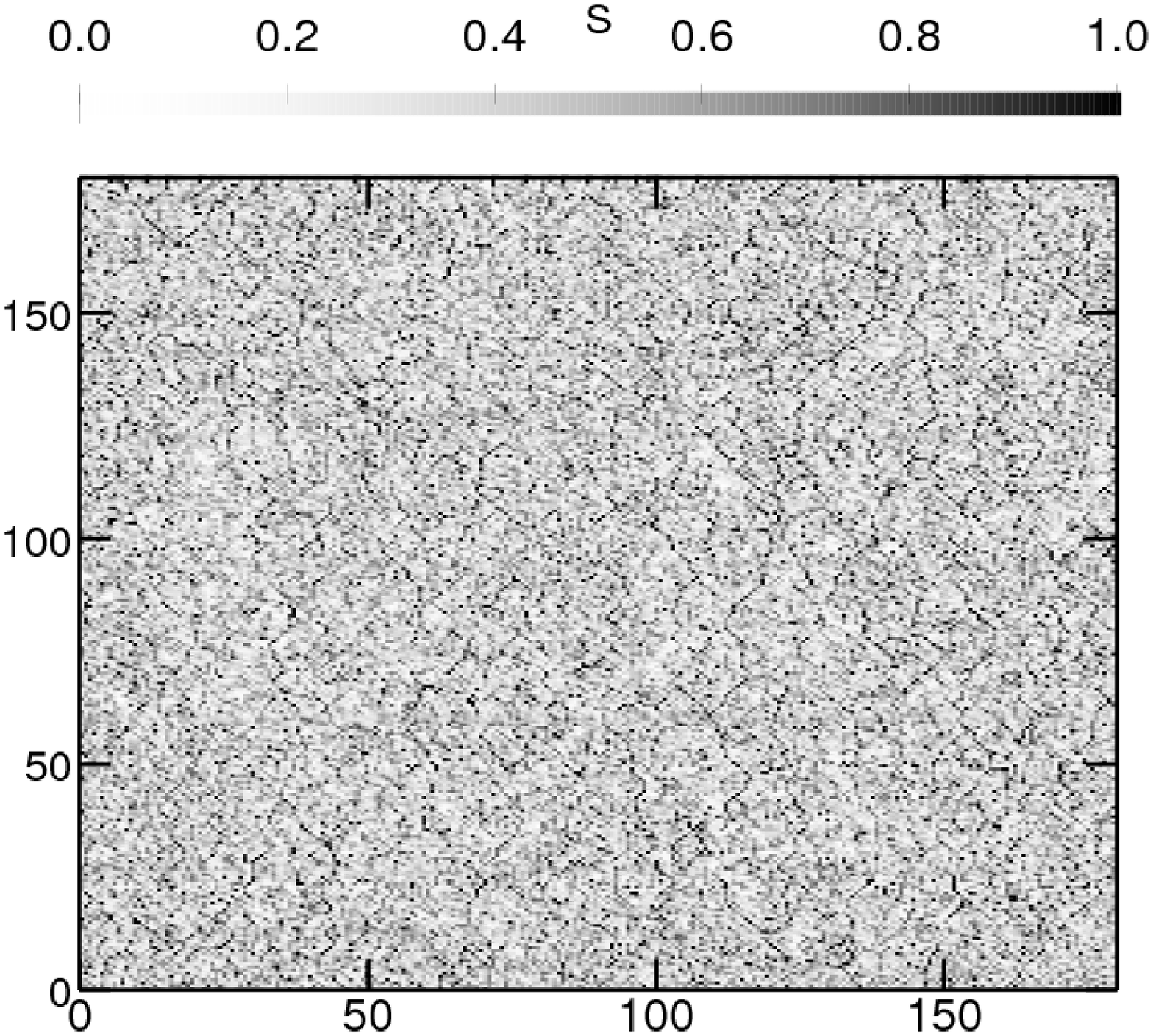}} ~
  \scalebox{0.32}{\includegraphics{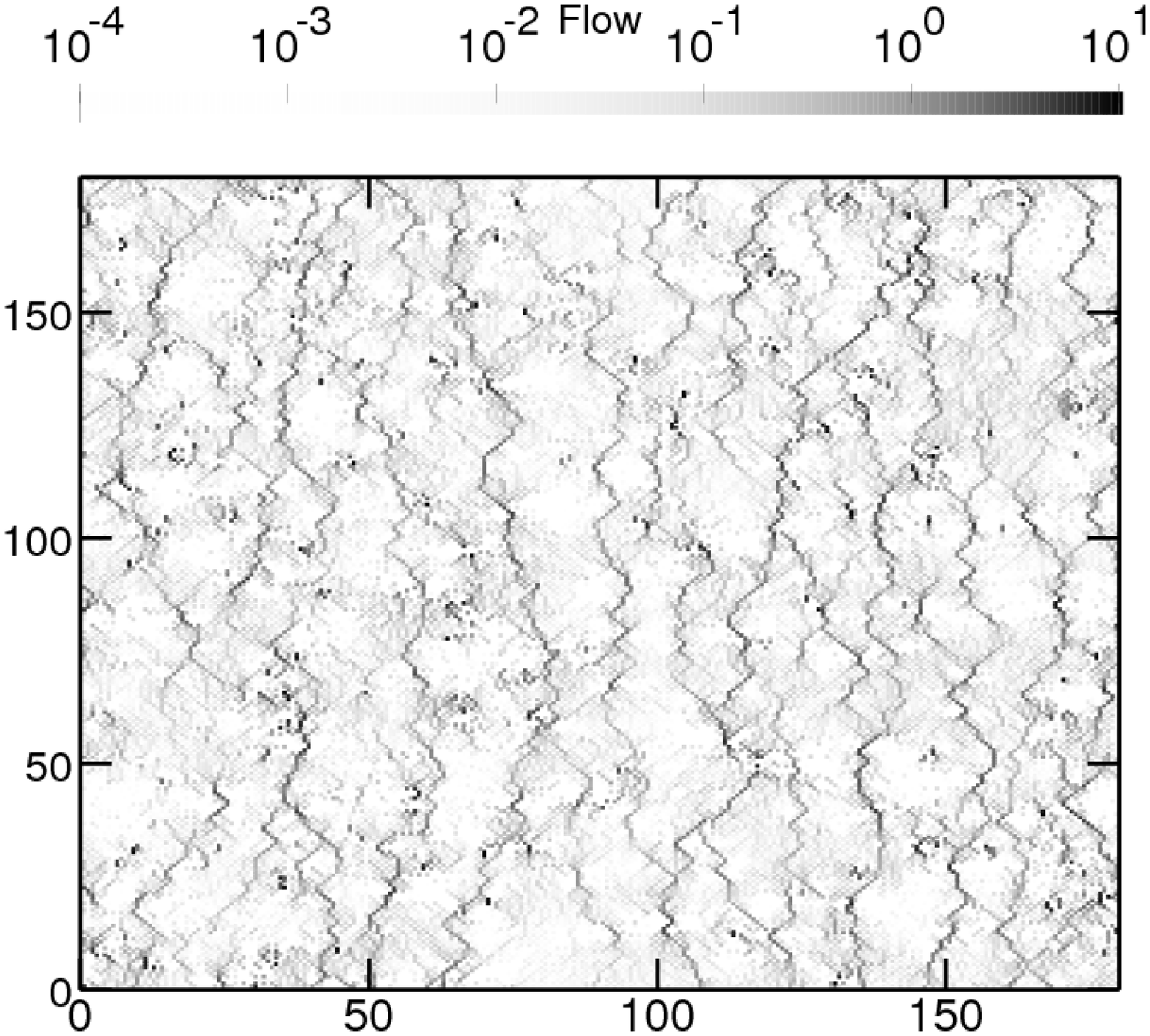}}
  \caption{
Snapshot of the spatial distribution of local saturations $s$ (left)
and absolute-valued flow-rates $|q|$ (right) under steady-state
conditions. The flow-rates are plotted on a logarithmic scale, in
arbitrary units. The overall direction of flow is up. Axes are
measured in lattice units ($\ell$). Bi-periodic boundary conditions
are used. $\M = 1$, $\Ca = 10^{-3}$ and $S = 0.5$.
\label{fig:flowdist}
}
\end{center}
\end{figure*}

An impression of the model is provided by Fig.~\ref{fig:flowdist}.
Connected strands of large non-wetting saturations can be seen (dark)
amidst patches of high wetting saturations (light). Internal
flow-rates span several orders of magnitude. Channels of both high and
low flow-rates are clearly visible. There are channels within
channels, forming a complex pattern. For the choice of parameters and
initialization shown in Fig.~\ref{fig:flowdist}, channels are dynamic.
At any given time flow-rates outside fast channels are relatively
slow, but when menisci block old channels the blocks are by-passed by
new channels, and no spatial region remains stagnant indefinitely.

\subsection{Definitions}

The main parameter of the model is the relative strength of capillary and
viscous forces, parametrized by the capillary number
\begin{equation}
\label{eq:ca}
  \Ca = \frac{\mueff Q}{\sigma \phi A} .
\end{equation}
Larger values of $\Ca$ correspond to faster flow, or lower surface tension.

The viscosity ratio $\M$ is defined as
\begin{equation}
  \M = \frac{\mu_{\rm nw}}{\mu_{\rm w}} ,
\end{equation}
where subscripts denote the non-wetting and wetting phases.
In this work a viscosity ratio $\M = 1$ (viscosity matched fluids) and
$\M = 10^{-4}$ (gas/liquid) as in the experiment has been explored.

The effective permeability $\keff$ is defined by
\begin{equation}
  \keff = \frac{Q}{Q_0} .
\end{equation}
which gives
\begin{equation}
\label{eq:effdarcy}
  Q = \keff Q_0 = -\frac{\keff \kappa_0}{\mueff} \frac{\phi A}{L_y} \Delta P .
\end{equation}
Eq.~\ref{eq:effdarcy} resembles Darcy's law, with $\keff \kappa_0$
substituted for the absolute permeability. It should be kept in mind
that $\keff$ depends on $Q$ and $S$, so Eq.~\ref{eq:effdarcy} should
not be interpreted as a linear relation between $Q$ and $\Delta P$.

A power law dependence between effective permeability and capillary number
is expressed as
\begin{equation}
\label{eq:gamma}
  \keff \sim \Ca^{\gamma} ,
\end{equation}
which is meant to be valid within some range for $\Ca$.

Given Eq.~\ref{eq:gamma} and considering that $Q \sim \Ca$
when all other parameters are fixed, the scaling of $\Delta P$
can be expressed as
\begin{equation}
\label{eq:pscaling}
  \Delta P \sim \keff^{-1} Q \sim \Ca^{1-\gamma} ,
\end{equation}
which together with the definition in Eq.~\ref{eq:beta} gives the
relation $\gamma + \beta = 1$.
Eqs.~\ref{eq:beta} and \ref{eq:pscaling} are valid when $Q \sim \Ca$,
i.e., all parameters in $\Ca$ except $Q$ are kept constant.
This is the case in the Oslo and Bozeman experiments. Eq.~\ref{eq:gamma}
relates two dimensionless quantities and is valid regardless of the manner by
which $\Ca$ is changed.


\subsection{Initialization and steady-state}

The reached state is not necessarily independent of how the initial conditions
are chosen -- different initial distributions of the two phases may lead to
different states. For this reason it is important to compare different
initialization schemes.

In order to achieve a desired saturation the throats are initially filled with
only one phase, after which the other phase is inserted in randomly chosen
throats until the desired saturation is reached. Two schemes for inserting a
phase into a throat are considered here. First, the chosen throat may be filled
completely. Second, a bubble of some small, fixed volume and consisting of two
menisci may be inserted at some random position within the throat. Several
bubbles may then exist in a single throat. If they overlap, they are merged. In
the first scheme there are no capillary pressure drops initially, as these
vanish at the ends of the throats. In the second scheme the total capillary
pressure drop across the system averages out to zero, but there are finite,
random capillary pressure drops on the scale of the pore throats. These two
schemes will be referred to as large bubble and small bubble initialization,
respectively.

After the desired saturation is reached, two schemes are considered for the
approach to steady-state. First, the simulation may be run with a surface
tension which is initially low, and then gradually increased until the desired
$\Ca$ is reached. Second, the simulation may be run with constant parameters
throughout. This will be referred to as gradual and abrupt initialization,
respectively.

After a sufficient number of timesteps the system always settles into a dynamic
steady-state, unless a spanning cluster of one of the phases is formed, in
which case the spanning cluster carries all flow and the system ceases to
fluctuate. This will be referred to as the {\it spanning cluster transition}.
The dynamic steady-state is characterized by persistent fluctuations around
average values with no perceivable drift. It is observed that the range of
parameters which lead to a dynamic steady-state coincides with those which
gives flow of both phases.


\section{Results}


\subsection{Two states}

It is found that there are two possible states, one state with stagnant
clusters and one without.

The stagnant cluster state is reached by large bubble and abrupt
initialization, for a narrow range of saturations $S \approx 0.2$. The stagnant
clusters are spatial regions with slow flow-rates and slowly changing
saturation distributions. The locations of the stagnant clusters are static
throughout the simulations. The stagnant clusters form after only a very small
number of timesteps and their locations depend on the initial distribution of
phases. In this sense, the stagnant cluster state is a state with memory of
initial conditions. The meaning of this is discussed in Sec.~\ref{sec:meta}.
The range of saturations which lead to the stagnant cluster state is affected
by the choice of merging scheme and coalescence parameter.

Reaching the stagnant cluster state requires particular choices for
initialization scheme and parameters. By contrast, the state without stagnant
clusters is reached by all initialization schemes and parameters, except those
that produce the stagnant cluster state or a spanning cluster. No memory of
initial conditions is observed. There are structures: channels of flow and
connected strands of the non-wetting phase, but these do not appear to have any
particular length-scale. All throats and menisci appear to affect and be
affected by the overall flow. This is in contrast with the stagnant cluster
state, where the configurations within stagnant clusters appear to be insulated
from the overall flow pattern. Because all menisci participate in forming the
pattern of flow, the state without stagnant clusters will be referred to as the
foam state.

Starting from the stagnant cluster state, the stagnant clusters will mobilize
if $\Ca$ is increased above some threshold. If $\Ca$ is then returned to its
original value, the foam state results. The stagnant cluster state can thus
transition to the foam state given a perturbation. The opposite transition is
never observed. This leads to identifying the stagnant cluster state as
metastable. The foam state, on the other hand, is robust with regards to
changes in $\Ca$ -- after a period of increased or decreased $\Ca$ the system
returns to the same state.


\subsection{Power law exponents}

\begin{figure}[t]
\begin{center}
  \scalebox{0.60}{\includegraphics{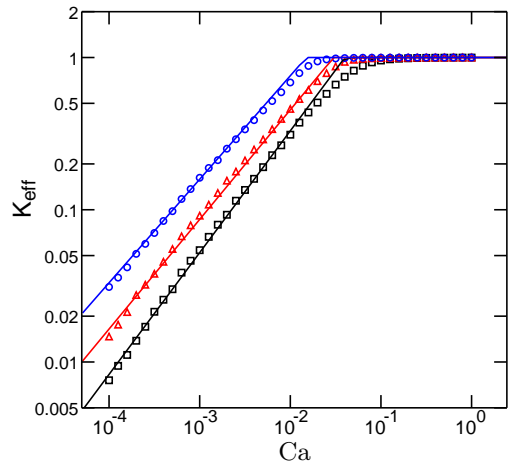}}
  \caption{
(Color online)
Scaling of $\keff$ with $\Ca$ for saturations 0.1 (circles), 0.2
(triangles) and 0.4 (squares). Straight lines are power law fits below
a threshold and equal to 1 above it. Error bars are omitted as they
are smaller than the point representations. $\M = 1$. System size is
32x64.
\label{fig:foamscaling}
}
\end{center}
\end{figure}

Fig.~\ref{fig:foamscaling} shows simulation results for $\keff$ in the
foam state, for a broad range of $\Ca$ and $\M = 1$. Three values of
$S$ are shown. Four simulation runs underlie each data point. Each run
has independent realizations of the network disorder and initial phase
distribution. The simulation is ended when the initial transient is
judged to occupy no more than the initial one-third of the run;
$\keff$ is then averaged over the last third. The variation of the
averaged $\keff$ between independent runs is quite small. If the
variations were plotted as error bars on Fig.~\ref{fig:foamscaling},
they would be smaller than the point representations.

A power law is fitted by eye, with a transition at some upper
threshold value for $\Ca$, around which the system has a smooth but
distinct transition. The threshold value is denoted $\Caup$. The fit
for $\keff$ is given by
\begin{equation}
\label{eq:powerlaw}
  \keff =
    \begin{dcases}
      \left( \frac{\Ca}{\Caup} \right) ^{\gamma} & \Ca < \Caup , \\
      1 & \Ca > \Caup .
    \end{dcases}
\end{equation}

From Fig.~\ref{fig:foamscaling} a distinct bending is observed with
respect to the fit. A possible cause of the bending may be the
numerical error caused by the Euler timestepping method, which becomes
more serious at low $\Ca$. The bending may also signal some cross-over
behavior between two regimes with different scaling exponents. Also,
if the power law is governed by some critical point at a finite $\Ca$,
omitting this will cause an apparent bending near the critical point.
A further discussion of a possible modification of the power law is
given in Sec.~\ref{sec:foam}.

\begin{figure}[t]
\begin{center}
  \scalebox{0.60}{\includegraphics{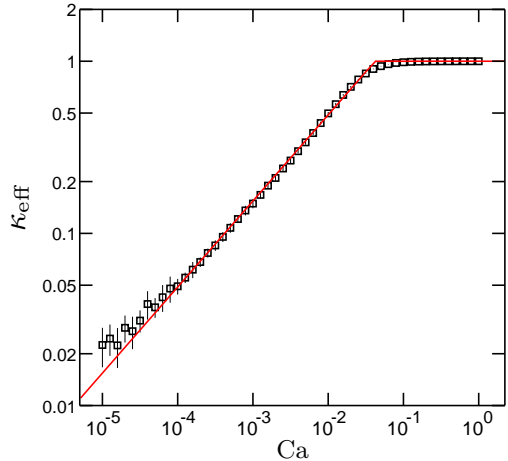}}
  \caption{
(Color online)
Scaling of $\keff$ with $\Ca$ for $S = 0.2$ in the stagnant
cluster state. Below the threshold $\Caup$, a power law with
$\gamma = 0.50$ is shown. Error bars are included, as these
become larger than the point representations for $\Ca < 10^{-4}$.
$\M = 1$. System size is 32x64.
\label{fig:stagnantscaling}
}
\end{center}
\end{figure}

Fig.~\ref{fig:stagnantscaling} shows simulation results for
$\keff$ in the stagnant cluster state. Some bending exists in the
limit of very small $\Ca$. It is in this limit that error bars begin
to become substantial. For a range of values of $\Ca$ exceeding two
decades, good power law behavior is observed.

Tab.~\ref{tab:gamma} reports the best fit values for $\gamma$ and
$\Caup$ for the foam and stagnant cluster states, for $\M = 1$ and $\M
= 10^{-4}$ and various values of $S$. In the foam state the best fit
for the exponent $\gamma$ depends on $S$, and varies in the range of
$0.67 - 0.80$.

At the large $\Ca$ limit of the law ($\Ca \gtrsim \Caup$) the
permeability approaches the single-phase permeability of the system,
$\keff = 1$. The existence of this {\it viscous transition} is not
surprising; it indicates the point where capillary pressure drops
become perturbations of a dominating field of viscous pressure drops.
The transition has been investigated for hysteresis, which has not
been found. Its smooth, reversible behavior is reminiscent of a second
order phase transition. For $\Ca > \Caup$, $F=S$ for all values of
$S$. A precise estimation of $\Caup$ has not been attempted. The
results presented in Tab.~\ref{tab:gamma} are those resulting from the
fit according to Eq.~\ref{eq:powerlaw}.

The power law also has a limit for low $\Ca$. When $\Ca$ is decreased
below some threshold $\Calo$, a spanning cluster forms. At this {\it
spanning cluster transition} the saturation distribution and flow
pattern becomes static. Decreasing $\Ca$ further after the formation
of a spanning cluster does not change $\keff$. Independent runs at
$\Ca$ beneath the spanning cluster transition display significant
hysteresis, and the concept of a reproducible state is lost. Due to
the hysteresis effects, the spanning cluster transition is reminiscent
of a first order phase transition. A further discussion of this is
given in \cite{knudsen2006two}.

$\Calo$ depends on $S$. For the purpose of this work, the transition
is avoided by choosing $S$ in the range of $0.1 - 0.4$. For larger or
smaller values of $S$ the spanning cluster transition would appear
within the range of $\Ca$ shown in Fig.~\ref{fig:foamscaling}.

\begin{table}
\begin{center}
  \caption{
The scaling exponent $\gamma$ and threshold value $\Caup$
for different values of $\M$ and $S$. System size is 32x64.
\label{tab:gamma}
}
  \begin{tabular}[t]{p{0.2cm}cp{0.5cm}cp{0.5cm}cp{0.5cm}c}
  \\
  & $\M$ & & $S$ & & $\gamma$ & & $\Caup$ \\
  \cline{1-8} \noalign{\smallskip}
  \multicolumn{7}{l}{
  {\it Stagnant cluster state:}}\\
  & 1 & & 0.2 & & 0.50 & & 0.042 \\
  & $10^{-4}$ & & 0.2 & & 0.54 & & 0.040 \\
  \multicolumn{7}{l}{
  {\it Foam state:}}\\
  & 1 & & 0.1 & & 0.68 & & 0.015 \\
  & 1 & & 0.2 & & 0.72 & & 0.030 \\
  & 1 & & 0.3 & & 0.76 & & 0.035 \\
  & 1 & & 0.4 & & 0.80 & & 0.040 \\
  & $10^{-4}$ & & 0.1 & & 0.67 & & 0.017 \\
  & $10^{-4}$ & & 0.2 & & 0.71 & & 0.029 \\
  & $10^{-4}$ & & 0.3 & & 0.75 & & 0.037 \\
  & $10^{-4}$ & & 0.4 & & 0.78 & & 0.040 \\
  \end{tabular}
\end{center}
\end{table}


No size-dependence has been found for $\keff$. In addition to the
results for 32x64 systems in Tab.~\ref{tab:gamma}, identical values
have been found for 16x32 systems. Also, several representative points
have been checked for larger systems, up to 256x512, with
indistinguishable results.


\section{Discussion}


\subsection{The viscosity ratio}

It is somewhat surprising that $\keff$ has almost no dependence on the
viscosity ratio $\M$. The simulations display a strong dependence on
$\M$ whenever transients are considered, in agreement with established
experimental results. To the author's knowledge, no experiment has
studied a dependence on $\M$ under steady-state conditions.

The reason for the lack of sensitivity to $\M$ shown by the
simulations may be attributed to the lack of large single-phase
clusters under steady-state conditions. If the distribution of menisci
causes the system to resemble a foam, no flow of one phase is possible
without causing the other phase to flow in nearly equal measure. Then,
only the volume-weighted effective viscosity $\mueff$ matters; the
viscosity ratio $\M$ does not.

In the high $\Ca$ limit this explanation seems plausible. It is in
this limit that the picture of a foam with small variations in the
saturation distribution is most appropriate. However, when $\Ca$ is
reduced the distribution of saturation becomes more heterogeneous, and
a large viscosity contrast could be expected to become more important.
The absence of an increased $\M$ dependence at low $\Ca$ indicates
that the relative importance of capillary pressure drops relative to
viscous pressure drops, which increases with decreasing $\Ca$,
out-weighs the increased importance of $\M$.

In other words, the scaling of $\keff$ with $\Ca$ is dominated by the
effects of capillary pressure drops, with only a modest contribution
from heterogeneities in the viscosity distribution.

The viscosity ratio has a significant importance in at least one
respect. The spanning cluster transition at $\Calo$ is sensitive to
$\M$. For $\M \gg 1$ (more-viscous non-wetting fluid), $\Calo$ is
between $10^{-2}$ and $10^{-3}$ for all values of $S$. This does not
allow for a power law spanning much more than a single decade. For
this reason the case of $\M \gg 1$ is not considered in this work.


\subsection{The metastable state \label{sec:meta}}

How can the stagnant cluster state be claimed to be a state, given
that it only appears for a particular initialization scheme and the
location of stagnant clusters depends on the initial distribution of
phases? While the state retains memory of the initial distribution of
phases for a long time, possibly indefinitely, the macroscopic
averages are consistent from realization to realization. The
fluctuations of effective permeability within a single run and the
fluctuations of run averages between independent runs are of the same
order. This is true both when the individual runs are on identical
networks with different initial phase distributions and when the
individual runs have different realizations of the network disorder.
The variation of macroscopic averages between different simulation
runs are all consistent with these fluctuations. In this sense, the
macroscopic averages are not sensitive to the initial phase
distribution.

In other words, the stagnant cluster state has well-defined properties
despite displaying hysteresis. For this reason, it is referred to as a
state.

It should be noted that a stagnant cluster does not consist of a
single phase. Due to the way the simulations are initialized, both
phases will in general exist within the cluster. Therefore, there will
be some capillary pressure drops within. These pressure drops do not
seem to be organized, they are random and cancel out. This indicates
that the perimeter of the stagnant cluster is formed early in the
simulation, before the menisci within the cluster organize themselves
in response to the flow.

If $S$ is larger than 0.3 the stagnant clusters are always mobilized,
eventually resulting in the foam state, even for the case of large
bubble and abrupt initialization. This is due to the larger pressure
drop required to obtain the desired flow-rate at these saturations.
The larger pressure drop prevents stagnant clusters from forming, and
the foam state results. For a discussion on the dependence of $\Delta
P$ on $S$, see \cite{knudsen2002relation}.


\subsection{The scaling of the foam state \label{sec:foam}}

A recent paper \cite{sinha2012effective} explores a power law which is
different from Eq.~\ref{eq:gamma}. The simulator used in their paper
is similar to the one used here. They explore a power law on the form
$(\Delta P - \Delta P_c) \sim \Ca^{\beta}$ and find $\beta \approx
0.5$. $\Delta P_c$ is a threshold pressure drop, beneath which the
authors of \cite{sinha2012effective} state that no flow occurs. The
introduction of this new power law is in part motivated by the
wandering value of $\beta$ reported in \cite{grova2011two}.


The primary purpose of this work is to consider the power law given by
Eq.~\ref{eq:gamma}. It corresponds to the power laws reported by the
Oslo and Bozeman experiments. However, the bending observed in
Fig.~\ref{fig:foamscaling} does indicate that a power law of a
different form should also be considered.

In order to explore this possibility, a more general power law is
considered in the following, on the form
\begin{equation}
\label{eq:gammaone}
  (\keff - \kcr) \sim (\Ca - \Cacr)^{\gamma_1} ,
\end{equation}
where $ \Calo < \Ca < \Caup$. $\Cacr$ and $\kcr \equiv \keff(\Cacr)$
may now be interpreted as critical values for the control and order
parameter, respectively. In the following they will be treated as
independent parameters in an optimization problem, aiming to obtain
the best possible fit to the power law.

Part of the motivation for considering Eq.~\ref{eq:gammaone} is to see
whether a dependence of $\Cacr$ on $S$ can remove the apparent
dependence of the exponent on $S$. If an $S$-dependent critical value
is ignored this can be misinterpreted as an $S$-dependent exponent
\footnote{A.~Hansen, private communication.}.

\begin{figure}[t]
\begin{center}
  \scalebox{0.60}{\includegraphics{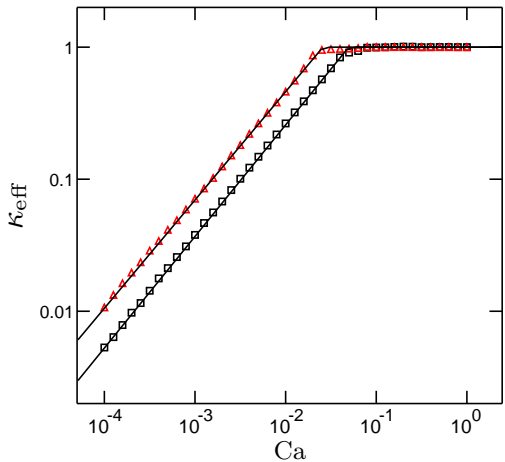}}
  \caption{
(Color online)
Scaling of $\keff$ with $\Ca$ for saturations 0.2 (triangles) and 0.4
(squares). The porous medium was initialized without disorder.
Otherwise the simulations are identical to those in
Fig.~\ref{fig:foamscaling}. $\M = 1$. $S = 0.2$ has $\gamma = 0.82$
and $\Caup = 0.025$. $S = 0.4$ has $\gamma = 0.85$ and $\Caup =
0.048$.
\label{fig:nodisorder}
}
\end{center}
\end{figure}

In addition to the simulation results reported for the foam state in
Tab.~\ref{tab:gamma}, simulations without disorder are considered
here. Fig.~\ref{fig:nodisorder} shows results from simulations with
$\M=1$, $S=0.2$ and $S=0.4$, identical to those reported in
Fig.~\ref{fig:foamscaling} except that all throat radii are set equal
to $r = 0.2 \ell$. With no disorder, the exponent $\gamma$ increases
to $0.82$ and $0.85$ and less bending is observed. As before, the
power laws have been fitted by eye.

The critical values $\Cacr$ and $\kcr$ are determined by minimizing
the error of the best fit for the power law. Instead of fitting by
eye, the best fit is now found using a least squares method,
considering only the data points within the interval $10^{-4} < \Ca <
10^{-2}$. The error is minimized with repsect to variations of the two
critical values by means of simulated annealing.

\begin{figure}[t]
\begin{center}
  \scalebox{0.57}{\includegraphics{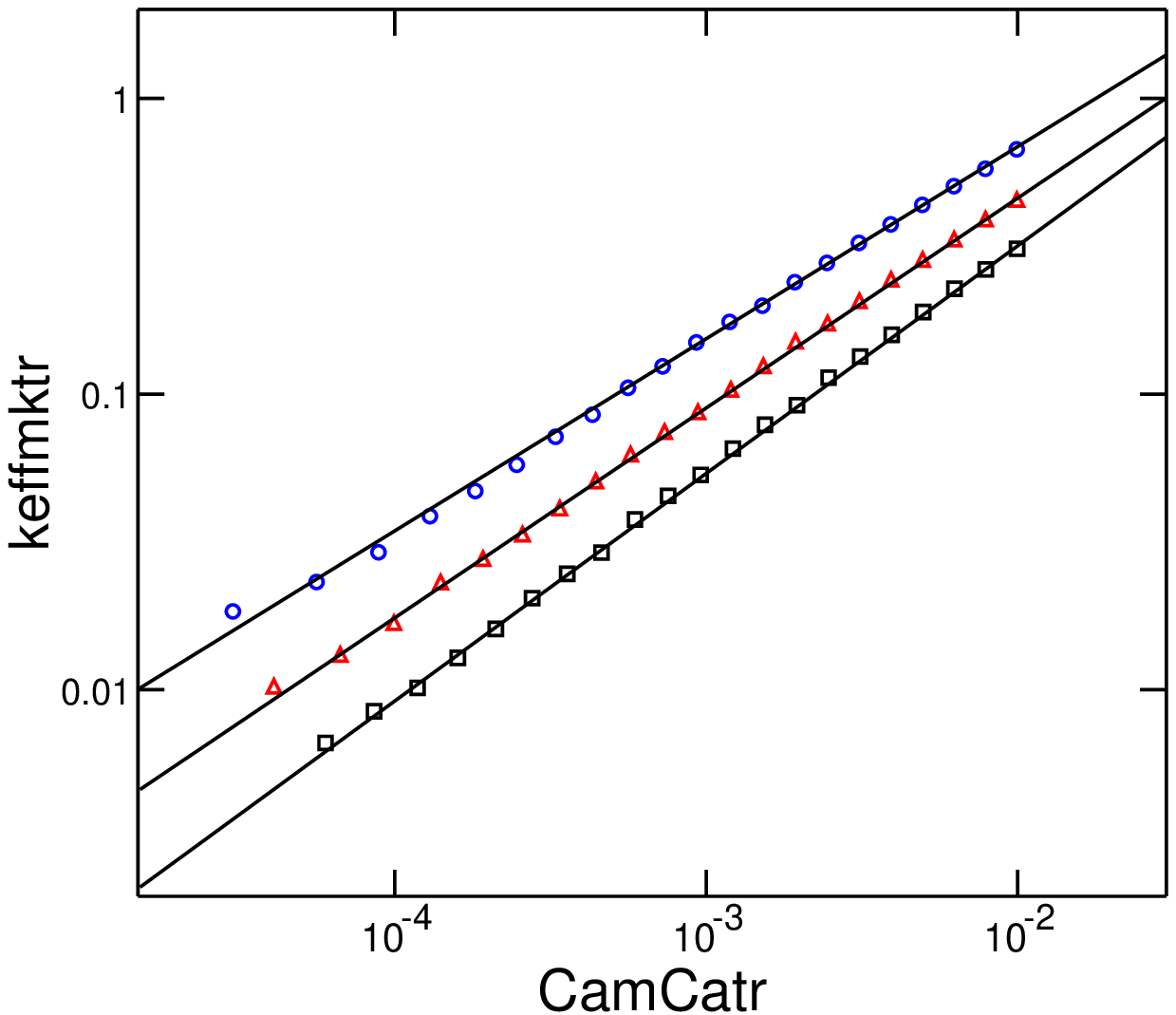}}
  \caption{
(Color online)
Scaling of $\keff-\kcr$ with $\Ca-\Cacr$ for saturations 0.1
(circles), 0.2 (triangles) and 0.4 (squares). Straight lines are power
law fits. $\M = 1$.
\label{fig:straightened}
}
\end{center}
\end{figure}

Tab.~\ref{tab:straightened} reports the results from this procedure.
Fig.~\ref{fig:straightened} displays the results and power law fits
for the same values of $\M$ and $S$ as Fig.~\ref{fig:foamscaling}.

The values found for $\Cacr$ are beneath the range of $\Ca$ which are
accessible to the simulator at this time. Therefore, the results
reported in Tab.~\ref{tab:straightened} may not be reliable, as they
are extrapolated by an assumption of a power law on the form of
Eq.~\ref{eq:gammaone} and can not be verified independently of this.

For the case of no disorder, the datasets give reasonable fits to a
power law without the introduction of critical values. When an attempt
is made to estimate the critical values, they are found to be orders
of magnitude smaller than in the disorderd case. This makes
extrapolating their values highly inaccurate, therefore they have
simply been set to zero.

The conclusion that can be drawn from this analysis, is that a power
law on the form of Eq.~\ref{eq:gammaone} can explain the observed
bending with respect to Eq.~\ref{eq:gamma}. This does not remove the
dependence of the exponent on $S$. The estimated values for $\Cacr$
are of the same order of magnitude as $\Calo$. Hysteresis and
numerical inaccuracies prevent the hypothesis $\Calo = \Cacr$ from
being explored further. It has also been found that the exponent is
sensitive to disorder, as are the hypothesized critical values.
Further work should investigate whether this remains true in three
dimensions.

\begin{table}
\begin{center}
  \caption{
The scaling exponent $\gamma_1$, $\Cacr$ and $\kcr$ for different
values of $\M$ and $S$, with and without disorder. Only the foam state
is considered.
\label{tab:straightened}
}
  \begin{tabular}[t]{p{0.2cm}cp{0.5cm}cp{0.5cm}cp{0.5cm}cp{0.5cm}c}
  \\
  & $\M$ & & $S$ & & $\gamma_1$ & & $\Cacr$ & & $\kcr$ \\
  \cline{1-10} \noalign{\smallskip}
  \multicolumn{9}{l}{
  {\it Disorder:}}\\
  & 1 & & 0.1 & & 0.65 & & $7.0 \cdot 10^{-5}$ & & $1.3 \cdot 10^{-2}$ \\
  & 1 & & 0.2 & & 0.71 & & $5.9 \cdot 10^{-5}$ & & $4.4 \cdot 10^{-3}$ \\
  & 1 & & 0.3 & & 0.75 & & $6.1 \cdot 10^{-5}$ & & $3.1 \cdot 10^{-3}$ \\
  & 1 & & 0.4 & & 0.77 & & $4.0 \cdot 10^{-5}$ & & $9.8 \cdot 10^{-4}$ \\
  & $10^{-4}$ & & 0.1 & & 0.66 & & $7.7 \cdot 10^{-5}$ & & $1.4 \cdot 10^{-2}$ \\
  & $10^{-4}$ & & 0.2 & & 0.71 & & $6.1 \cdot 10^{-5}$ & & $5.1 \cdot 10^{-3}$ \\
  & $10^{-4}$ & & 0.3 & & 0.75 & & $5.2 \cdot 10^{-5}$ & & $2.4 \cdot 10^{-3}$ \\
  & $10^{-4}$ & & 0.4 & & 0.77 & & $3.9 \cdot 10^{-5}$ & & $1.6 \cdot 10^{-3}$ \\
  \multicolumn{9}{l}{
  {\it No disorder:}}\\
  & 1 & & 0.2 & & 0.81 & & 0.0 & & 0.0 \\
  & 1 & & 0.4 & & 0.85 & & 0.0 & & 0.0 \\
  \end{tabular}
\end{center}
\end{table}


\subsection{Comparison with experiments -- $\keff(\Ca)$}

The Oslo experiment \cite{tallakstad2009prl, tallakstad2009pre} found
an exponent $\beta = 0.54 \pm 0.08$, corresponding to $\gamma = 0.46
\pm 0.07$. Their result covers roughly two decades in $\Ca$. Their
definition of $\Ca$ is different from Eq.~\ref{eq:ca}; by the
definition used here their experiments are carried out at $\Ca$
roughly between $3 \cdot 10^{-6}$ and $2 \cdot 10^{-4}$.

They use a two-dimensional Hele-Shaw cell: a monolayer of glass beads
sandwiched between transparent plates. At the inlet 15 evenly spaced
syringes inject the two phases with equal flow-rate through each
syringe. The two phases are injected through every other syringe, with
7 syringes injecting air, fixing fractional flow to $F = 7/15$. The
other syringes inject a water-glycerol solution which will simply be
referred to as water in the following.

Initial conditions consist of a fully water-saturated medium.
Initially, pockets of air grow at the injection points. Eventually,
the pockets become detached from the injection points and travel
through the medium until they escape at the far end. It is observed
that smaller pockets are more likely to be trapped. The pockets of air
are embedded within a spanning cluster of water. In the steady-state
no drift is observed in the average global pressure drop, and the
distribution of air pockets appears to be translationally invariant in
a statistical sense.

The Oslo experiment explores values of $\Ca$ and $S$ which,
interpreted in terms of the simulations, places them below the
spanning cluster transition. In the simulations only one of the phases
flow after $\Ca$ is reduced below $\Calo$. In the Oslo experiment both
phases flow despite the existence of a spanning cluster, due to the
simultaneous injection. Also, below the spanning cluster transition,
hysteresis is expected. This suggests that the experiment only sees
simultaneous flow of air and water due to forcing air to flow by the
conditions at the inlets. This will in the following be referred to as
forced fractional flow, which should be distinguished from the case of
free fractional flow. Free fractional flow is the case in the
simulations here when $\Ca > \Calo$.

There are substantial differences between the Oslo experiment and the
simulations reported here. However, the Oslo experiment shares some
similarities with the stagnant cluster state. First, the stagnant
clusters are characterized by low, nearly vanishing flow-rates
surrounded by channels of fast flow, held in place by a perimeter of
capillary pressure drops. The trapped pockets of air in the Oslo
experiment must be held in place by a similar mechanism. Second, the
stagnant clusters are formed by initial conditions with few capillary
pressure drops and develop until they reach a size where they are
by-passed by flow without being mobilized. The pockets of air in the
Oslo experiment are formed at the inlets, where they are allowed to
grow until they reach a size where flow of the surrounding water
forces them to either mobilize or fragment. In both cases, clusters
with a characteristic size are produced.

Could the setup of the Oslo experiment obtain the foam state? Two
approaches may be suggested. First, large values of $\Ca$ could be
explored, possibly obtained by adding surfactants to reduce surface
tension. This would also allow an investigation of the transition at
$\Caup$. Second, instead of injecting the two phases at alternating
syringes, the two phases could be mixed in tee junctions prior to
injection through the syringes. This would prevent the build-up of
large air pockets at the inlets, possibly resulting in a different
state.

In \cite{tallakstad2009prl, tallakstad2009pre} a scaling argument is
presented for $\beta = \gamma = 0.5$. A close relative of this
argument is presented in Sec.~\ref{sec:scaling}, where the comparison
between the Oslo experiment and the stagnant cluster state is
revisited.

The Bozeman experiment \cite{rassi2011nuclear} found an exponent
$\beta$ in the range of $0.30-0.45$, corresponding to $\gamma$ in the
range of $0.55-0.70$. Their result covers a little more than a single
decade in $\Ca$. Their definition of $\Ca$ is similar to the one used
here, and is roughly in the range between $10^{-4}$ and $10^{-3}$.
Their definition of $\Ca$ uses the viscosity of water rather than the
volume-weighted average of the two phases, however the water
saturation does not change dramatically except for at the largest
$\Ca$. The different definitions do not seriously affect the
exponents.

They use a three-dimensional bead pack, where water floods the sample
at gradually increasing flow-rates after an initial preparation with
simultaneous flow of air and water. The initial preparation results in
a partially saturated state. As the sample is flooded by water, air
escapes until the only pockets of air left within the sample are
immobile. Even at the largest flow-rates the pattern of flow is found
to be different from that of the fully water saturated sample, meaning
that there are always some pockets of air remaining. These pockets are
suggested to be quite small, as they do not show up on MR images.
Repeated floodings of a single sample gives comparable results,
suggesting that the method produces a reproducible state.

The partially saturated state is achieved by a procedure which mixes
the two phases before they are injected, as opposed to the Oslo
experiment. This could prevent the formation of stagnant clusters. At
the end of the initial preparation only water flows. The saturation
decreases gradually until the spanning cluster transition is reached,
at which all air is trapped. When the flow-rate is increased, some air
mobilizes temporarily until the spanning cluster transition is reached
again. If the experiment resembles the foam state as it approaches the
spanning cluster transition, the partially saturated state will have
an effective permeability {\it near} that of the foam state at similar
$S$ and $\Ca$.

The best estimate for $\beta$ provided in \cite{rassi2011nuclear} is
$0.33$. This gives $\gamma \approx 0.67$, which could suggest that
their experiment has indeed seen evidence of the foam state. However,
there are significant differences between the Bozeman experiment and
the simulations. The simulations are two-dimensional and saturation is
kept constant under steady-state conditions, as opposed to gradually
decreasing while the sample is flooded.  In Sec.~\ref{sec:more} it is
argued that the stagnant cluster exponent $\gamma \approx 0.5$ should
be valid for both two and three-dimensional porous media. This is
further indication the the Bozeman experiment does {\it not} produce
the stagnant cluster state, which leaves the foam state. It is not
known whether the foam state exponents change in three dimensions.


\subsection{Comparison with experiments -- $S(\Ca)$}

Both the Oslo and Bozeman experiments have reported decreasing $S$
with increasing $\Ca$.  In both cases it is found that the low $\Ca$
limit is $S \approx 0.5$.

The Bozeman experiment finds that $S$ strongly decreases for $\Ca
\gtrsim 10^{-3}$. Because water is the only injected phase it is not
surprising that $S$ decreases for larger $\Ca$, this agrees with the
spanning cluster transition moving to lower $S$ for larger $\Ca$. This
also suggests that the Bozeman experiment transitions into the regime
of free fractional flow. If the sharp decrease is caused by a
transition from the regime of forced to free fractional flow, the
sharp decrease can be explained as the disappearance of hysteresis
effects.

In the Oslo experiment the decrease of $S$ is harder to explain. As
$\Ca$ is increased, the characteristic size of trapped air pockets
decreases. The number of air pockets must somehow be governed by the
mechanisms which govern flow of air. More air pockets should allow
more air to flow. The decrease of $S$ with increasing $\Ca$ suggests
that the number of air pockets needed to maintain a fixed $F$ does not
grow fast enough to compensate the reduced size of the air pockets.

At the $\Caup$ transition seen in the simulations, fractional flow
curves have a distinct transition from a non-trivial dependence on $S$
and $\Ca$, to $F=S$ for all values of $S$ and all values of $\Ca >
\Caup$. This means that if the Oslo experiment were to increase $\Ca$
further, they should at some point observe a reversal in the $S(\Ca)$
trend. $S$ should increase to an ultimate value of $S=F=7/15$ above
the $\Caup$ transition. The value of $\Ca$ which gives a minimum value
of $S$ most likely coincides with the transition from forced to free
fractional flow.


\section{Scaling argument for $\beta = \gamma = 1/2$ \label{sec:scaling}}

In \cite{tallakstad2009prl, tallakstad2009pre} an argument is
presented for why $\beta = \gamma = 1/2$. It will be referred to as
the Oslo argument in the following.

An argument which is closely related to the Oslo argument is given in
Secs.~\ref{sec:postulates} and \ref{sec:expon}. The two arguments lead
to the same scaling exponent. Secs.~\ref{sec:osloarg} and
\ref{sec:more} compare the two arguments and discuss their
applicability.

\subsection{Postulates \label{sec:postulates}}

Three postulates are stated. Given that these are justified,
$\beta = \gamma = 1/2$ results.

The physical picture is that of two phases somehow distributed through and
injected into a two-dimensional porous medium, initially with few capillary
pressure drops. Eventually menisci with large capillary pressure drops
develop, causing some regions to resist flow. The imposed flow is then carried
in regions that remain open to flow. These two regions are the stagnant cluster
volume and the dissipative volume, respectively. The stagnant clusters are held
in place by a perimeter of capillary pressure drops and are as large as they
can be without mobilizing. The dissipative volume consists of a connected
network of channels spanning the system.

{\it Postulate 1.}
The volume $V$ of the sample can be divided into two sub-volumes,
\begin{equation}
  V = \Vdiss + V_{\rm stag} ,
\end{equation}
where the dissipative volume $\Vdiss$ satisfies
\begin{equation}
\label{eq:dissvolume}
  D \approx \Ddiss \equiv \sum_{\rm diss} d \sim
  \sum_{\rm diss} u|\nabla P| \sim \Vdiss u^2 .
\end{equation}
$D$ is the total dissipation within the sample, $d$ is the dissipation
within a throat, the sums run over the throats within the dissipative
volume, $u$ is the mean velocity within the dissipative volume and
$\nabla P$ is the mean pressure gradient within this volume.
Separating $d$ into the product of the mean values of $u$ and $\nabla
P$ relies on flow within the dissipative volume respecting Darcy's law
and being reasonably homogeneous, or fluctuations not being
correlated.

{\it Postulate 2.}
Consider the balance of pressure gradients along the stagnant cluster
perimeters. Everywhere along the perimeter there is a pressure balance $p_a -
p_b \approx p_c$, where $p_a$ and $p_b$ are the pressures on either side of the
cluster perimeter and $p_c$ is the pressure drop caused by the meniscus on the
perimeter. $p_c$ scales with the surface tension $\sigma$, which gives
\begin{equation}
\label{eq:capbalance}
  \nabla P \cdot l^* \propto \sigma ,
\end{equation}
where $l^*$ is the length of the path that connects the extreme points of the
stagnant cluster, i.e., from the point of highest to the point of lowest
surrounding pressure. $l^*$ is a characteristic length scale of the stagnant
clusters, and should be comparable with both their mean width $l_x$ and mean
length $l_y$.

{\it Postulate 3.}
It is necessary to relate $l^*$ to $\Vdiss$. Assume that the mean width of the
flowing channels does not scale with $\Ca$. Both experiments and simulations
observe that channels are only a single pore-width wide, supporting the
assumption. Also, assume that the aspect ratio of the stagnant clusters
$l_y/l_x$ does not scale with $\Ca$, such that
\begin{equation}
  l^* \approx (l_x+l_y) \sim l_x \sim l_y .
\end{equation}
Smaller $\Ca$ gives larger clusters, but the aspect ratio does not necessarily
change. In experiments, clusters that mobilize are observed to be elongated
along the direction of flow, while the smaller, stagnant clusters have $l_x
\approx l_y$. In the simulations, the stagnant clusters appear slightly
elongated in the direction of flow, and larger clusters seem to be more
elongated. No $\Ca$ dependence is observed.

The dissipative volume scales only with the number of channels when their width
is constant. The number of channels $N_x^{\rm ch}$ scales as
\begin{equation}
N_x^{\rm ch} \sim L_x/l_x \sim l_x^{-1} ,
\end{equation}
where $L_x$ is the width of the sample. This gives
\begin{equation}
\label{eq:volbalance}
  l^* \sim \Vdiss^{-1} .
\end{equation}
The tortuosity of the flow channels $\tau$ also plays a role in the argument.
It can be defined as the ratio of the mean channel length to the length of the
sample $L_y$. Given the assumptions above, the scaling of $\tau$ with $\Ca$ is 
\begin{equation}
  \tau \approx \frac{\sum_i l^*_i}{L_y} 
  \sim N_y^{\rm cl} l^* \sim (l^*)^{-1} l^* \sim {\rm const} ,
\end{equation}
where $N_y^{\rm cl} \equiv L_y/l_y$ is the number of clusters over which the sum runs.
Larger clusters give larger, but fewer twists and turns for the flow channels, which
results in a tortuosity $\tau$ which does not scale with $\Ca$.

\subsection{The scaling exponents \label{sec:expon}}

Consider the effective cross-section $\Sigma_{\rm eff} = \Vdiss / L_y$. The
flow velocity must scale as
\begin{equation}
\label{eq:effcross}
  u \sim \frac{Q}{\Sigma_{\rm eff}} \sim \frac{Q}{\Vdiss} .
\end{equation}
Next, consider the scaling of the flow velocity with the pressure gradient,
\begin{equation}
\label{eq:gradient}
  u \sim \nabla P \sim \frac{\Delta P}{\tau L_y} .
\end{equation}
Combining Eqs.~\ref{eq:capbalance}, \ref{eq:volbalance}, \ref{eq:effcross}
and \ref{eq:gradient} gives
\begin{equation}
  \frac{Q}{\Vdiss} \sim u \sim \nabla P \sim \frac{\sigma}{l^*} \sim \sigma \Vdiss ,
\end{equation}
which leads to
\begin{equation}
  \Vdiss^2 \sim \frac{Q}{\sigma} \sim \Ca .
\end{equation}
Using Eq.~\ref{eq:dissvolume} gives an expression for the total dissipation as
\begin{equation}
\label{eq:dissca}
  D \sim \Vdiss u^2 \sim \Vdiss \left( \frac{Q}{\Vdiss} \right)^2
  \sim \frac{Q^2}{\Vdiss} \sim \frac{Q^2}{\Ca^{1/2}} .
\end{equation}
The dissipation can also be expressed by using Eq.~\ref{eq:effdarcy},
\begin{equation}
\label{eq:disskeff}
  D = \Delta P \cdot Q \sim \frac{Q^2}{\keff} ,
\end{equation}
from which the scaling is obtained as
\begin{equation}
\label{eq:scaling}
  \keff \sim \Ca^{1/2} \Rightarrow \gamma = 1/2 .
\end{equation}

Other scaling results that can be read out of the preceding equations are
\begin{equation}
  \Delta P \sim \nabla P \sim (l^*)^{-1} \sim \Ca^{1/2} \Rightarrow \beta = 1/2 .
\end{equation}
In the last scaling laws, dimensional quantities are related to a dimensionless
number.  It should be understood that this is done for $Q \sim \Ca$, which is
only valid when all other dimensional quantities are kept constant.
Eq.~\ref{eq:scaling} is valid regardless.

\subsection{Comparison with the Oslo argument \label{sec:osloarg}}

The Oslo argument is very similar to the argument given here. The
assumptions of postulates 1, 2 and 3 are made by both. However, there
are three significant differences between the arguments.

First, Eq.~16 of \cite{tallakstad2009pre} gives some length scale
$l^*$ (not identically defined as the $l^*$ used here, but similar) as
being proportional to the ratio of capillary pressure drop to overall
viscous pressure drop. This is similar to postulate 2. The argument
then ignores the capillary pressure drop, giving $l^* \propto \Delta
P^{-1}$. However, the capillary pressure drop should not be ignored
since the interest is in scaling with $\Ca$, which contains the
surface tension. Since the Oslo argument uses $\Ca \sim Q$ their
conclusion is not affected by this, however it reduces the generality
of the argument, which is best expressed in terms of dimensionless
quantities.

Second, Eqs.~16 and 17 of \cite{tallakstad2009pre} contain the ratio
$\Delta P / L$, which is meant to be the pressure gradient surrounding
a stagnant cluster. Taking $L$ as the system length $L_y$ ignores the
tortuosity of flow paths as they wrap around stagnant clusters. This
tortuosity could in principle scale with $\Ca$. However, the
assumptions of postulate 3, which the Oslo argument also make, mean
that $\tau$ does not scale with $\Ca$. This is an important point
within the scaling argument and should not be ignored. If $\tau$
scaled with $\Ca$, the conclusion $\Delta P \sim \Ca^{1/2}$ would no
longer be valid.

Third, the Oslo argument considers the case where the phase contained
in stagnant clusters has a very low viscosity, and presents their
argument as a limiting case. The simulations suggest that the
viscosity ratio is of little importance. The argument as presented
here also does not rely on a particular viscosity ratio. If there is a
non-unity viscosity ratio, $\Ca$ should be defined with the
volume-weighted effective viscosity of the dissipative volume rather
than the total volume, otherwise the argument is unaffected by $\M$.
Any discrepancy between the simulations and the argument due to
different definitions of $\Ca$ does not affect the scaling exponents,
because the saturation within the dissipative volume does not scale
with $\Ca$. The same should be true for the experiment.

\subsection{More on the scaling argument \label{sec:more}}

In the Oslo experiment the largest air pockets migrate while small ones are
trapped. The migration of a large air pocket could lead to fragmentation. Small
pockets of air could be released into the channels of flowing water, leaving
behind a now trapped air pocket. The small, flowing pockets could exit the
system, or coalesce with other trapped air pockets, causing them to migrate
until they fragment, such that the process repeats itself. This suggests the
mechanism behind forced fractional flow: over-sized pockets of air produced at
the inlets begin to migrate, then fragment, leading to a cascade of
coalescence, migration and fragmentation through the system.

The migrating cluster state is a very different picture from that of the
stagnant cluster state, where no migration is seen. There are clearly
substantial differences between the simulations and the Oslo experiment.
However, the origin of the $\beta = \gamma = 1/2$ exponent appears to be the
same in both, namely the applicability of postulates 1, 2 and 3.

The argument is not meant to be exact. In the case of the Oslo experiment, the
elongation of the largest air pockets is likely to be connected with the
mechanism which allows fractional flow. Because the aspect ratio of stagnant
clusters is important to postulate 3 the argument is likely to be modified by
an improved understanding of this mechanism. The separation into two volumes,
where one contains all dissipation and obeys Darcy's law and the other is
perfectly balanced by a perimeter of capillary pressure drops, is also an
approximation. The stagnant clusters are not perfectly non-dissipative in the
simulations, and the air pockets in the Oslo experiment are not all completely
trapped.

An advantage of simulations over experiments is that simulations have absolute
knowledge of the system configuration. This allows tests of the scaling
argument to be carried out, by testing its intermediate steps. It was found
that $\Vdiss / V \sim \sqrt{\Ca}$ and $u \sim \sqrt{Q} \sim \sqrt{\Ca}$. In
order to obtain $\Vdiss$ and $u$ it is necessary to determine a cut-off $\dtr$
for throat dissipations, with throats having $d > \dtr$ belonging to the
dissipative volume. $\dtr$ is determined by increasing it to the point where
the total dissipation contained above the threshold begins to decrease rapidly.

\begin{figure}[t]
\begin{center}
  \scalebox{0.60}{\includegraphics{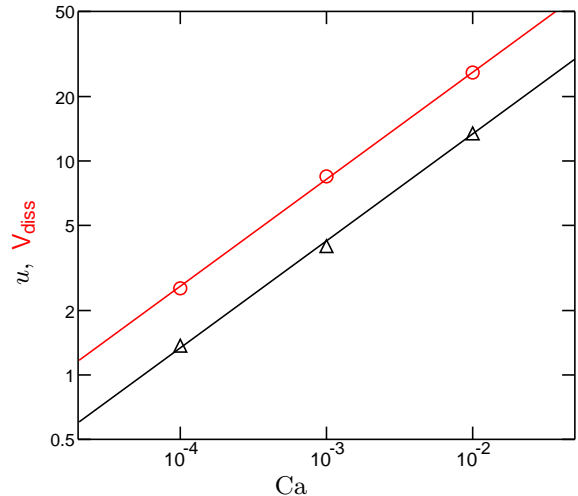}}
  \caption{
(Color online)
Scaling of mean velocity $u$ (triangles) and dissipative volume $\Vdiss$
(circles) with $\Ca$. Power laws (straight lines) are added with exponents 0.5.
The units on the y-axis refers to the percentage of dissipative to total
volume; velocities have arbitrary units. All simulations are within the
stagnant cluster state, with $S = 0.20$ and $\M = 1$.
\label{fig:scaling}
}
\end{center}
\end{figure}

The values for $u$ and $\Vdiss$ obtained in this way are plotted for three
values of $\Ca$ in Fig.~\ref{fig:scaling}. They scale as they are supposed to,
in support of the scaling argument.


In three dimensions, the perimeter of a stagnant cluster is a two-dimensional
surface and the dissipative volume consists of sheets of flow surrounding these
perimeter surfaces. Postulates 1 and 2 do not need to be modified by this.
Postulate 3 changes to the following: the sheets of flow have a thickness which
does not scale with $\Ca$ and the stagnant clusters obey the scaling $l^* \sim l_x
\sim l_y \sim l_z$. Given these modifications the rest of the scaling argument
survives unaltered for three dimensions. The result $\beta = \gamma = 1/2$ is
therefore expected to be valid for both two and three-dimensional systems.


An argument for a foam state exponent is not likely to share any
similarities with the argument for the stagnant cluster state. None of
the three postulates apply to the foam state. There is no separation
into two distinct volumes of low and high flow-rates, as can be seen
from inspecting histograms of flow-rates from the two states. No
regions appear to be stable with respect to either saturation
distribution or flow-rates -- flowing channels fluctuate across the
entire system -- which means that a balance criterion similar to
postulate 2 is not applicable. The absence of stable structures makes
it difficult to search for a characteristic length scale, while the
power law distribution of flow-rates suggest that no characteristic
length scale exists.

\section{Conclusion}

Network simulations of two-phase flow in porous media have been used
to investigate the scaling of $\keff$ with $\Ca$ under steady-state
conditions with fixed saturation. Two values of $\M$ were used,
corresponding to viscosity matched fluids and gas/liquid. Two states
were identified and named: the stagnant cluster state and the foam
state.

The metastable stagnant cluster state scales with $\gamma = 0.50$ and
$0.54$ for viscosity matched fluids and gas/liquid respectively. An
argument in support of $\gamma = 1/2$ has been given. The argument
applies to both the stagnant cluster state and the Oslo experiment.
More work is needed to explore the scaling in the foam state.

The Oslo experiment has been interpreted in terms of the distinction
between forced and free fractional flow. Their low values of $\Ca$ and
observed dependence between $S$ and $\Ca$ is indication that they
observe forced fractional flow, which indicates that their system is
governed by the boundary conditions at the inlets. The Bozeman
experiment, on the other hand, may have observed the foam state.

Upper and lower bounds have been established for the power law. At
$\Ca < \Calo$ a spanning cluster transition prevents power law scaling
for the case of free fractional flow, though the Oslo experiment
indicates that scaling persists under conditions of forced fractional
flow. At $\Ca > \Caup$ a viscous transition results in $\keff \approx
1$. The spanning cluster transition and viscous transition appear to
be first and second order, respectively.

Discussions with E.~G. Flekk{\o}y, A. Hansen, K.~J. M{\aa}l{\o}y, S.
Sinha and K.~T. Tallakstad are gratefully acknowledged. H.~A. Knudsen
is fondly remembered.

\end{document}